\documentclass[conference]{IEEEtran}

\usepackage{import}
\usepackage{amsthm}
\usepackage{threeparttable}
\usepackage{amsmath,amssymb,amsfonts}
\newtheorem{definition}{Definition} 
\usepackage{multirow}
\usepackage{enumitem}
\usepackage{xspace}
\usepackage{listings}
\usepackage[misc]{ifsym}
\usepackage{mfirstuc}
\usepackage{bbding}
\usepackage{fancyhdr} 
\usepackage{pgfplots}
\pgfplotsset{compat=1.12}
\usepackage{subfigure}
\usepackage{filecontents}
\usepackage{caption}
\usepackage{makecell}
\usepackage{ifthen}
\usepackage{academicons}
\usepackage[square,numbers,sort&compress]{natbib}
\usepackage[colorlinks,
            linkcolor=blue,       
            anchorcolor=blue,  
            citecolor=blue,        
            ]{hyperref}
\usepackage{multicol}
\usepackage[nameinlink]{cleveref}
\crefname{figure}{Figure}{Figures}
\crefname{subfigure}{Figure}{Figures}
\crefname{section}{Section}{Sections}
\crefname{subsection}{Section}{Sections}

\usepackage[ruled,linesnumbered]{algorithm2e} 

\SetAlFnt{\small}
\SetAlCapFnt{\small}
\SetAlCapNameFnt{\small}
\SetAlCapHSkip{0pt}
\IncMargin{-\parindent}
\SetKwInOut{Parameter}{Parameters}
\SetKwProg{Fn}{Function}{:}{}

\usepackage{soul}	
\setstcolor{red}

\long\def\com#1{}

\newboolean{showchange}
\setboolean{showchange}{true}

\ifthenelse{\boolean{showchange}}
{

}
{

}

\newcommand{\blind}[1]{}		

\newcommand{\mysect}[1]{{\vspace{0.2cm} \noindent\kb{#1}}}		
\newcommand{\kw}[1]{{\em#1}}	
\newcommand{\ksb}[1]{\textbf{{\footnotesize #1}}}	
\newcommand{\kb}[1]{{\textbf{#1}}}	
\newcommand{\kn}[1]{\texttt{\small #1}}	

\theoremstyle{definition}

\usepackage{hyperref}
\usepackage{amsmath,amssymb,amsfonts}
\usepackage{algorithmic}
\usepackage{graphicx}
\usepackage{textcomp}
\usepackage{xcolor}

\usepackage{svg}
\usepackage{booktabs}
\usepackage{caption}
\usepackage{subfigure} 

\def\BibTeX{{\rm B\kern-.05em{\sc i\kern-.025em b}\kern-.08em
    T\kern-.1667em\lower.7ex\hbox{E}\kern-.125emX}}

\lstdefinestyle{cypher}{
     basicstyle=\small\selectfont\ttfamily,
     breaklines=true, 
     keywordstyle=\bfseries\color{blue},
     columns=flexible,
     morekeywords={MATCH,return,RETURN,match,CREATE,VIEW,AS,CONSTRUCT,WITH,DELETE,UNION,WHERE}, 
     numbers=none, 
     frame=none,
     xleftmargin=0pt,
     escapeinside={(*@}{@*)},
}

\newcommand{\PG}{$\mathcal{PG}$\xspace}  
\newcommand{\PGS}{$\mathcal{PGS}$\xspace}  
\newcommand{\PatG}{$\mathcal{Q}$\xspace}  %
\newcommand{\OPatG}{$\mathcal{Q}_o$\xspace}  %
\newcommand{\VPatG}{$\mathcal{V}$\xspace} 
\newcommand{\VPatGs}{$\mathbb{V}$\xspace}
\newcommand{\SVPatGs}{$\mathbb{V'}$\xspace}
\newcommand{\VC}{$s_{vc}$\xspace} 
\newcommand{\VCS}{$\mathbb{S}_{vc}$\xspace} 
\newcommand{\VM}{$s_{vm}$\xspace} 
\newcommand{\MPatG}{$\mathcal{MP}$\xspace} 
\newcommand{\NF}{$\mathcal{N}$\xspace} 
\newcommand{\EF}{$\mathcal{E}$\xspace} 

\newcommand{\Labels}{\textbf{L}\xspace}
\newcommand{\Keys}{\textbf{P}\xspace}
\newcommand{\Values}{\textbf{V}\xspace}

\newcommand{\Powerset}[1]{$\mathcal{P}(#1)$\xspace}

\newcommand{\VMTSet}{$\mathcal{S}_\textsf{VMT}$\xspace}
\newcommand{\VMTMap}{$\mathcal{M}_\textsf{VMT}$\xspace}

\newcommand{\DBHit}[1]{$\textsf{DBHit}_\textsf{#1}$}
\newcommand{\squote}[1]{\lq #1\rq}
\newcommand{\nodepat}{{\kw{P}$_\textsf{N}$}}
\newcommand{\relinfo}{\kw{R}}
\newcommand{\pattern}{\kw{P}}
\newcommand{\patseg}{{\kw{P}$_\textsf{S}$}}
\newcommand{\relpat}{{\kw{P}$_\textsf{R}$}}
\newcommand{\mysys}{\textsc{Mv4pg}\xspace}  

\begin{document}
\title{MV4PG: Materialized Views for Property Graphs
\\
}

\author{
    \IEEEauthorblockN{Chaijun Xu$^{1}$, Xingdi Wei$^{1}$, Yu Zhang$^1$, Kaiwei Li$^2$, Xiaowei Zhu$^2$, Ke Huang$^2$, Tao Wang$^2$, Shipeng Qi$^3$}
    \IEEEauthorblockA{$^1$ University of Science and Technology of China, Hefei, China}
    \IEEEauthorblockA{$^2$ Tsinghua University, Beijing, China}
    \IEEEauthorblockA{$^3$ Huazhong University of Science and Technology, Wuhan, China}
    \IEEEauthorblockA{\{xuchaijun,wxd352541141\}@mail.ustc.edu.cn, yuzhang@ustc.edu.cn, \\ chnlkw@163.com, \{coolerzxw,wangtao.waves,qishipengqsp\}@gmail.com, 569078986@qq.com}
}
\maketitle

\begin{abstract}
Graph databases are getting more and more attention in the highly interconnected data domain, and the demand for efficient querying of big data is increasing.
We noticed that there are duplicate patterns in graph database queries, and the results of these patterns can be stored as materialized views first, which can speed up the query rate.
So we propose materialized views on property graphs, including three parts: view creation, view maintenance, and query optimization using views, and we propose for the first time an efficient templated view maintenance method for containing variable-length edges, which can be applied to multiple graph databases.

In order to verify the effect of materialized views, we prototype on TuGraph and experiment on both TuGraph and Neo4j. The experiment results show that our query optimization on read statements is much higher than the additional view maintenance cost brought by write statements. The speedup ratio of the whole workload reaches up to 28.71x, and the speedup ratio of a single query reaches up to nearly 100x.

\end{abstract}




\section{Introduction}
With the booming development of big data, graph databases are increasingly crucial in applications such as social network analysis, recommendation systems, financial fraud detection, and transaction analysis~\cite{fan2022big}. Research into further accelerating the query rate of graph databases has also attracted increasing amounts of people.

In the workload of graph databases, there are often some patterns that will be queried repeatedly. If the results of these patterns are stored in advance, the query rate can be greatly accelerated by directly querying the stored results when needed, such an idea is very similar to that of materialized views defined in relational databases~\cite{palpanas2002incremental,himanshu2006incremental,ke2003efficient,koch2014dbtoaster,larson2006efficient,chaudhuri1995optimizing,halevy2000theory,flesca2001rewriting,goldstein2001optimizing}.
However, certain types of queries in graph databases, such as reachability queries containing variable-length edges, receive little attention in the context of relational databases. As a result, the implementation of materialized views in graph databases is still a fresh area waiting to be developed.

Some research has been conducted on materialized views in graph databases such as~\cite{da2020kaskade,han2024implementation,fan2016answering}. These studies leverage materialized views to optimize query performance. However, when the graph database is updated, the consistency between the view and the original database might be broken, requiring maintenance of the view to restore consistency. Notably, these works do not address the maintenance of variable-length edges, which, as previously mentioned, play a crucial role in specific types of queries within graph databases and warrant more focused attention.

\mysect{Our proposed approach.} To address this issue, we propose an approach that consists of three key components: view creation, view maintenance, and query optimization using views.
In the view creation phase, we define the creation statements based on the syntax of GQL~\cite{ISO/IEC39075}, the standard graph database query language. This ensures compatibility and standardization across graph database systems.
In the view maintenance phase, the maintenance template is predefined at the time of view creation. When the graph data is updated, the template parameters are substituted to generate specific maintenance statements, ensuring that the view remains consistent with the underlying graph data. 
In the query optimization phase using views, materialized views are ranked by their optimization potential, matched to the query graph via subgraph matching, and integrated into the query to improve execution efficiency.

\begin{figure}[!t]
  \centering
    \begin{minipage}{.5\textwidth}
    \centering
    \subfigure[Pattern Graph of Example Query] 
	{
			\centering      
			\includegraphics[scale=0.35]{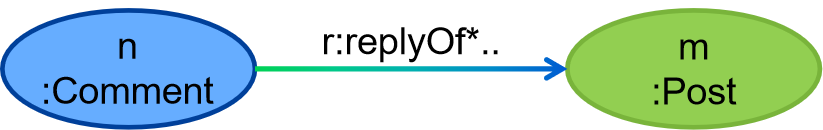}   
            \label{fig:example_view}
	}
      \subfigure[Pattern Graph of Example View] 
	{
			\centering      
			\includegraphics[scale=0.35]{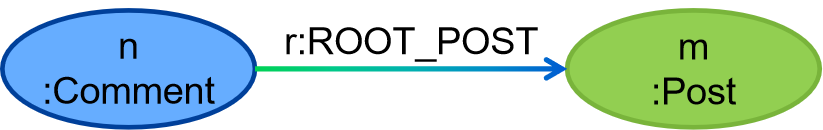}   
            \label{fig:example_view2}
	}
  \end{minipage}
  \caption{Example of View Creation}
  \label{fig:example_view_creation}
  \vspace{-2mm}
\end{figure}

Let’s assume a graph database is used to efficiently manage complex relationships and connections. The approach can be illustrated through the following three examples, each demonstrating one of the three key components:

\mysect{Example 1 (View Creation).} 
As shown in ~\cref{fig:example_view_creation}, to determine which \kn{Post} each \kn{Comment} belongs to, we query the graph database using the pattern graph depicted in ~\cref{fig:example_view}. This pattern graph includes variable-length edges (denoted by *..), where each edge represents a path with at least one hop and potentially infinite hops. However, querying uncapped variable-length edges can be time-consuming. 

To optimize this, we can store the query result as a new edge labeled \kn{ROOT\_POST}, as shown in ~\cref{fig:example_view2}. The subsequent sections on view maintenance and query optimization will build upon this materialized views.

\mysect{Example 2 (View Maintenance).} When the graph data is updated, such as deleting a \kn{Comment} or linking a new \kn{Comment} to a different \kn{Post}, the consistency of the materialized view and original graph data is disrupted, requiring the view to be maintained.

\begin{figure}
  \begin{minipage}{.5\textwidth}
    \centering
    \includegraphics[scale=0.35]{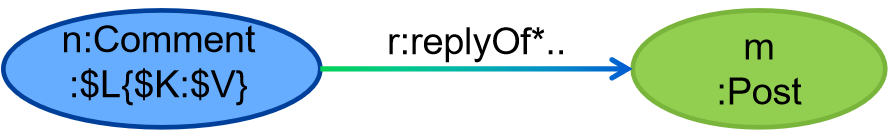}
  \end{minipage}
  
  \vspace{1mm}
  
  \begin{minipage}{.5\textwidth}
    \centering
    \includegraphics[scale=0.35]{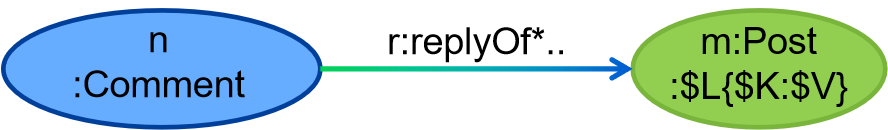}
  \end{minipage}

  \vspace{1mm}
  
  \begin{minipage}{.5\textwidth} 
    \centering
    \includegraphics[scale=0.35]{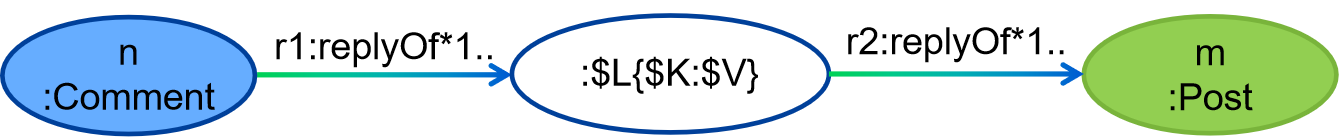}
  \end{minipage}
  \caption{Example of View Maintenance}
    \label{fig:example_maintenance}
\end{figure}

For efficient maintenance, only the affected views are incrementally updated, rather than being fully regenerated. 
For example, with the view in ~\cref{fig:example_view2}, if a node is deleted in the graph database, all impacted \kn{ROOT\_POST} edges must be removed. As shown in ~\cref{fig:example_view}, the deleted node could be at \kn{n} or \kn{m} or inside the variable-length edge \kn{r}. The three pattern graphs in ~\cref{fig:example_maintenance} are then used to identify the corresponding \kn{n}, \kn{m} node pairs. Here \$L, \$K, and \$V represent the label, primary key name, and primary key value, respectively. Based on these patterns, the corresponding \kn{n}, \kn{m} pairs are identified, and the affected \kn{ROOT\_POST} edges are deleted accordingly. 

To reduce the maintenance time, we can automatically generate a view maintenance template during view creation. When a node is deleted, we simply replace \$L, \$K, and \$V with the node's actual information, eliminating the need to regenerate the maintenance statement each time.

\mysect{Example 3 (Query Optimization Using Views).} Once the materialized view is in place, queries involving \kn{Comment-Post} relationships can be optimized. Instead of traversing variable-length edges in real-time, the query leverages the precomputed \kn{ROOT\_POST} edges. 

\begin{figure}[b]
  \centering
    \begin{minipage}{.5\textwidth}
    \centering
    \subfigure[Original Pattern Graph] 
	{
			\centering      
			\includegraphics[scale=0.35]{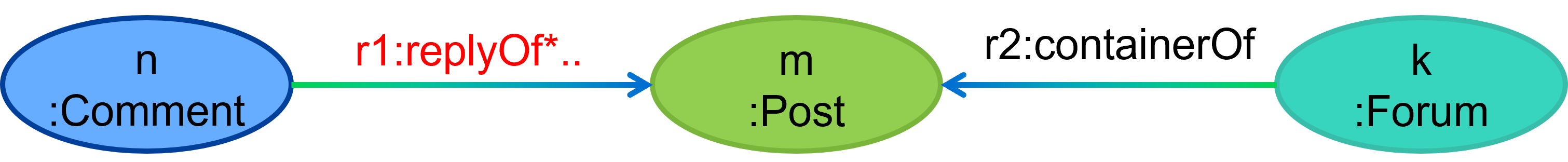}   
            \label{fig:example_optimization1}
	}
      \subfigure[Optimized Pattern Graph] 
	{
			\centering      
			\includegraphics[scale=0.35]{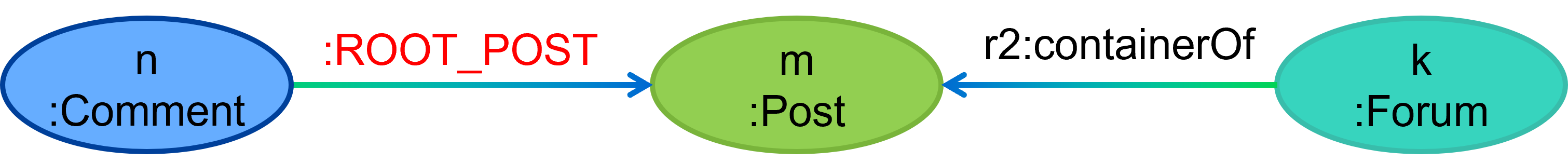}   
            \label{fig:example_optimization2}
	}
  \end{minipage}
  \caption{Example of Query Optimization Using Views}
  \label{fig:example_view_opt}
  \vspace{-2mm}
  \end{figure}

Given 
the query pattern graph in ~\cref{fig:example_optimization1}, we observe that the edge between \kn{n} and \kn{m} matches the view in \cref{fig:example_view2}.  As a result, this edge can be replaced with the view edge \kn{ROOT\_POST}, optimizing the query.

\mysect{Contributions.} In this paper, we propose an innovative solution for creating and maintaining materialized views on property graphs, with a focus on optimizing queries involving variable-length edges. A  central feature of our approach is the \mysys system, which seamlessly integrates view creation, maintenance, and query optimization into a unified framework. At its core is a templated view maintenance strategy that automatically generates maintenance templates during view creation. This method significantly reduces the overhead of regenerating maintenance statements after graph updates. By replacing only the relevant parts of the template with actual deletion information, we eliminate the need for redundant computations, ensuring both efficiency and scalability. Our solution is highly 
applicable to a wide range of property graph use cases, offering substantial performance improvements in query optimization.

The contributions of this paper are threefold:
\begin{itemize} 
    \item 
    We propose a view creation language that conforms to the GQL standard syntax,  facilitating easier adoption of our solution into existing graph query environments.
    \item 
    We introduce an efficient templated view maintenance approach for property graphs with variable-length edges.
    \item We demonstrate the practical effectiveness of the \mysys through a prototype implementation on TuGraph~\cite{TuGraph2023} (with code available on github~\cite{TugraphCode}), and comprehensive testing on both TuGraph and Neo4j~\cite{guia2017graph}. The results show that views significantly reduce the execution time of the overall workload, achieving a maximum speedup of 28.71×, with the highest speedup for a single statement approaching 100×. 
    Furthermore, the optimization brought by views far exceeds the additional maintenance cost associated with write statements.
\end{itemize}


\begin{table*}[]
\fontsize{8}{9.5}\selectfont
\caption{Comparison of View Implementation Features in Graph Databases}
\begin{center}
\begin{tabular}{c|c|c|c|c|c}
    \Xhline{1pt}
      & IVM4PGQ~\cite{szarnyas2018incremental} &  KASKADE~\cite{da2020kaskade} & MVS\&QP~\cite{pang2024materialized} & IS4VOPG~\cite{han2024implementation} & \mysys\\
    \Xhline{1pt}
    Inclusion of variable-length edges & Y & Y & Y & N & Y \\
    \hline
    View selection & N & A & A & M & M \\
    \hline
    View maintenance & Y & N & N & Y & Y \\
    \hline
    Query optimization & N & Y & Y & Y & Y \\
    \hline
    Multi-DBMS implementations & N & N & N & Y & Y \\
    \hline
    Experimental evaluation & N & Y & Y & Y & Y \\
    \hline
    \Xhline{1pt}
\end{tabular}
    \\\vspace{1em}
\parbox{0.85\linewidth}{\raggedright \textbf{Note}: \squote{Y} indicates that the feature is available, while \squote{N} means the feature is not available or not mentioned. In the view selection row, \squote{A} and \squote{M} represent automatic and manual, respectively.}\\
\label{tab:compare}
\end{center}
\end{table*}

\section{Preliminaries and Related Work}
\label{sec:pre}

\subsection{Preliminaries}
\mysect{Property Graph Model.} The property graph model, proposed by Rodriguez and Neubauer in 2010~\cite{rodriguez2010constructions}, utilizes nodes to represent entities, edges to denote relationships, and both nodes and edges can have their associated properties.

We adopt the definition of the property graph database model given by Angles~\cite{angles2018property}, and the schema is supported by Neo4j and TuGraph.
Assume that \Labels, \Keys, \Values represent the sets of labels, property names, and property values, respectively.
\begin{definition}[Property Graph~(\PG)]\label{def:property_graph}
    A property graph is defined as a tuple G = (N, E, $\rho$, $\lambda$, $\pi$) where:
\begin{itemize}
    \item $N$ is a finite set of nodes;
    \item $E$ is a finite set of edges such that $N \cap E=\emptyset $;
    \item $\rho: E \rightarrow (N\times N) $ is a total function mapping edges to ordered pairs of nodes;
    \item $\lambda : (N \cup E) \rightarrow$ \Powerset{\Labels} is a total function mapping nodes and edges to sets of labels (including the empty set), and \Powerset{\Labels} represents the power set of \Labels; 
    \item $\pi : (N \cup E)\times$ \Keys $\rightarrow$ \Powerset{\Values} is a partial function mapping properties of nodes/edges to sets of values.
\end{itemize}
\end{definition}

For $\lambda$, the set of labels associated with a single node or a single edge in this paper always contains exactly one label.

\mysect{Query Language.} The most popular query language for property graphs is Cypher~\cite{francis2018cypher}. 
Originating from Neo4j, it has been adopted by other vendors, including Amazon Neptune, Agens Graph, AnzoGraph, Katana Graph, Memgraph, RedisGraph, and SAP HANA~\cite{opencypher}.
To find all Comments posted under each Post, the Cypher query can be written as follows:

\begin{lstlisting}[style=cypher]
MATCH (n:Comment)-[r:replyOf*..]->(m:Post)
RETURN n,m
\end{lstlisting}
Here the \kn{MATCH} clause corresponds to the pattern graph of ~\cref{fig:example_view}.

\subsection{Related Work}
\label{sec:related_work}

\mysect{Materialized Views in RDBMS.} Materialized views have been studied in relational databases for several decades, and there is a great deal of work on them. ~\cite{palpanas2002incremental,himanshu2006incremental,ke2003efficient,koch2014dbtoaster,larson2006efficient} discusses the efficient maintenance of materialized views, including different kinds of views such as distributed and non-distributed aggregation functions, views containing join operators, etc.
~\cite{chaudhuri1995optimizing,halevy2000theory,flesca2001rewriting,goldstein2001optimizing} uses materialized views to reduce hits to the original database and thus optimize query efficiency, ~\cite{chaudhuri1995optimizing} discusses the timing of view optimization, suggesting that blindly using views may instead be more ineffective, ~\cite{goldstein2001optimizing} proposes a very efficient view-matching algorithm that can quickly find the right view to optimize query.

\mysect{Materialized Views in Property Graph DBMS.} 
~\autoref{tab:compare} shows related work on the materialized view of graph databases in recent years. 
~\cite{szarnyas2018incremental} first proposed incremental view maintenance in property graphs, converting graph query statements into the form of a relational algebra, using incremental maintenance algorithms in relational databases, but is more limited to theory and does not give evaluation results. ~\cite{da2020kaskade} automatically selects the required views for the workload and optimizes the query for rewriting using the views, focusing on the more specific queries in graph databases that contain variable-length edges, but not discussing the view maintenance part. ~\cite{pang2024materialized}, like ~\cite{da2020kaskade}, focuses only on view selection and query optimization on static graphs, and does not focus on the incremental maintenance component.
 ~\cite{han2024implementation} discusses materialized views in property graphs more systematically, including incremental maintenance and query optimization using views, and also uses a series of transformation rules to implement materialized views in both RDBMS and GDBMS, but does not take into account the more specific queries in graph databases that contain variable-length edges.
 In contrast, \mysys discusses both materialized view maintenance and query optimization using views, and gives a corresponding incremental maintenance approach for queries containing variable-length edges in graph databases, and the final experimental results in both Neo4j and TuGraph prove that \mysys plays a great role in query rate improvement.

\section{Overview of \mysys}
\begin{figure}[!t]
    \centering
    \includegraphics[width=\linewidth]{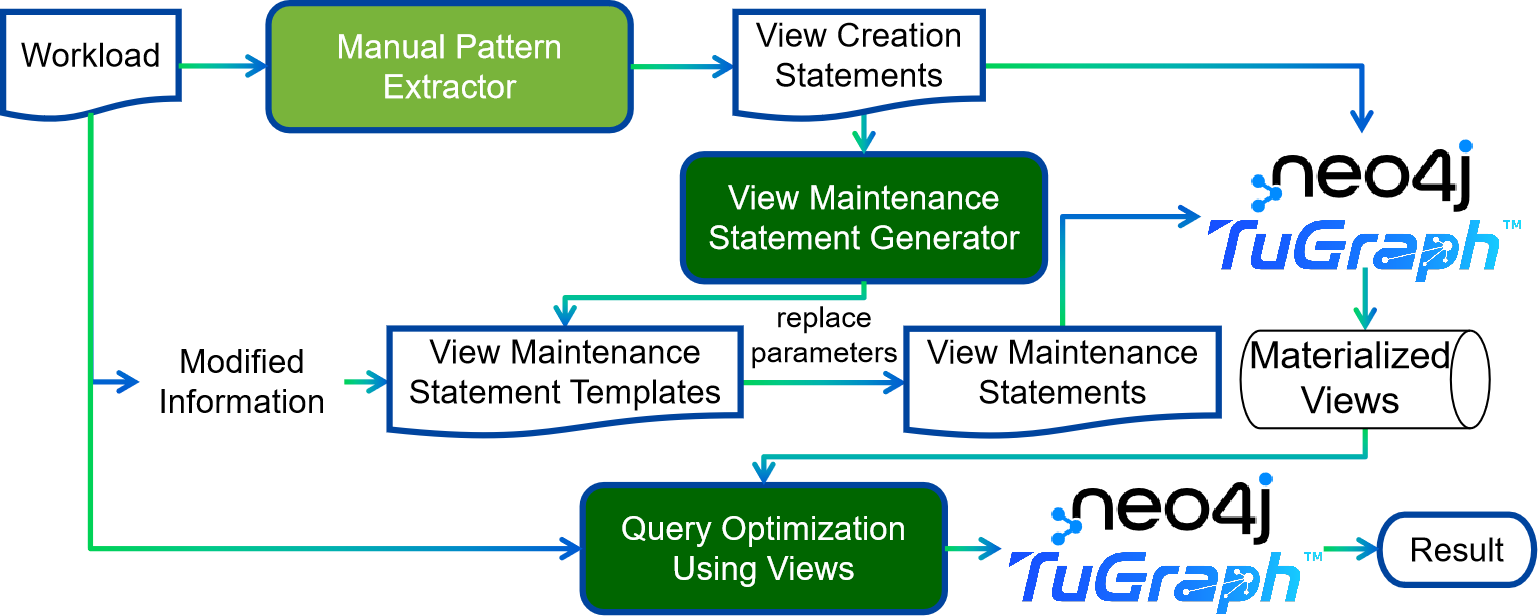}
    \caption{Overview of MV4PG}
    \label{fig:approach_overview}
\end{figure}
~\cref{fig:approach_overview} provides an overview of our \mysys system, designed to address the challenges of efficient view creation, maintenance, and query optimization in graph databases. For a given workload, we first manually extract common parts as views and materialize them in the graph database (Neo4j or TuGraph). Meanwhile, the {View Maintenance Statement Generator} creates {View Maintenance Statement Templates} based on the {View Creation Statements}. When modifications occur in the graph data as part of the workload, the \kw{modified information} is injected into the template parameters, generating the {View Maintenance Statements}, which are then executed for database maintenance. For the queries in the workload,  the {Query Optimization Using Views} module (see ~\cref{sec:query_opt1} for details) is applied to optimize them before execution in the graph database.

\autoref{tab:notations} provides a summary of the notations and their definitions used throughout this paper.
\begin{table}[!t]
\fontsize{8}{9.5}\selectfont
\caption{Notations}
\begin{center}
\begin{tabular}{c|c}
    \Xhline{1pt}
    \textbf{Notations} & \textbf{Definitions} \\
    \Xhline{1pt}
    \PatG, \OPatG & (optimized) pattern graph of a cypher query \\
    \hline
    
    \hline
    \VPatG & pattern graph of a view \\
    \hline
    \VPatGs & the set of \VPatG \\
    \hline
    \SVPatGs & sorted \VPatGs \\
    \hline 
    \VC & view creation statement \\
    \hline
    \VCS & all view creation statements in the graph \\
    \hline
    \NF & node information including labels and properties \\
    \hline 
    \multirow{2}{*}{\EF} & edge information including \NF of source node, \\
     & \NF of end node, labels and properties of the edge itself \\
    \hline
    \MPatG & path in \PatG that matches the view \\
    \hline
    \Xhline{1pt}
\end{tabular}
\label{tab:notations}
\end{center}

\end{table}

\section{View Creation And Maintenance}
\label{sec:view_creation_and_maintenance}

\subsection{View Creation}

As shown in ~\cref{fig:approach_overview}, the public patterns of the query statements are manually extracted from the workload, and the corresponding view creation statements are given, which are directly inputted into the graph database system. At the same time, for each view creation statement, we will generate the corresponding view maintenance statement template, which is put into ~\cref{sec:maintenance} to talk about it in detail.

Since existing graph database systems do not support the materialized view, we use part of the GQL grammar to extend the Cypher grammar, the relevant core grammar is shown in ~\autoref{fig:grammar_view_creation}.

\begin{figure}
    \centering\small
\begin{tabular}{llll}
(View) &
$v$&::=& \kn{CREATE VIEW} VName \kn{AS} \\ 
&&&\squote{(} \kn{CONSTRUCT} \pattern{}~ \kn{MATCH} \pattern{} \squote{)}; 
\\
(PatElem) & 
\pattern{} &::=& ( \nodepat{} ( \patseg{} )* ) ;\\
(PatSeg)& 
\patseg{} &::=& \relpat~ \nodepat;\\
(NodePat)& 
\nodepat{} &::=& \squote{(} ( Var )? ( NodeLabels )? ( Prop )? \squote{)} ; \\
(RelPat) &
\relpat{} &::=& ( \squote{\texttt{<-}} \relinfo{} ? \squote{\texttt{->}} ) \\
&& & $\mid$ ( \squote{\texttt{<-}} \ \relinfo{} ?\ \squote{\texttt{-}} ) \\
&& &$\mid$ ( \squote{\texttt{-}} \ \relinfo{} ?\ \squote{\texttt{->}} ) \\
&& &$\mid$ ( \squote{\texttt{-}}\ \relinfo{} ?\ \squote{\texttt{-}} ) \\
&& &; \\
(RelInfo) &
\relinfo &::=&\squote{[} ( Var )? ( RelTypes )? Range? \\
&&           &\quad ( Prop )?  \squote{]} ;
\end{tabular}
    \caption{Syntax of View Creation}
    \label{fig:grammar_view_creation}
\end{figure}

The main part of our view is the pattern element (PatElem) of \kn{Match} clause that represents a path from the start to the end. The PatElem of the \kn{Construct} clause has only two nodes and one view edge, the two nodes are the source and the end of the PatElem of the \kn{Match} clause. The role of the view is to compress a path that may contain variable-length edges into a single view edge, speeding up the query rate.

\mysect{Example.} A concrete example of the view definition is as follows:
\begin{lstlisting}[style=cypher]
CREATE VIEW ROOT_POST AS (
    CONSTRUCT (c)-[r:ROOT_POST]->(p)
    MATCH (c:Comment)-[:replyOf*..]->(p:Post) 
)
\end{lstlisting}
This statement creates a view called \kn{ROOT\_POST}, which identifies all (\kn{c},\kn{p}) pairs where a node \kn{c} of type \kn{Comment} connects via a variable-length edge to a node \kn{p} of type \kn{Post}. For each such pair, an edge of type \kn{ROOT\_POST} is created with \kn{c} pointing to \kn{p}.
The \kn{replyOf*..} notation, specifically \kn{replyOf*1.. $\infty$}, indicates that the edge has a variable length, with a minimum length of 1 and no upper limit ($\infty$).

\subsection{View Maintenance}
\label{sec:maintenance}

\begin{figure}
    \centering
    \includegraphics[width=\linewidth]{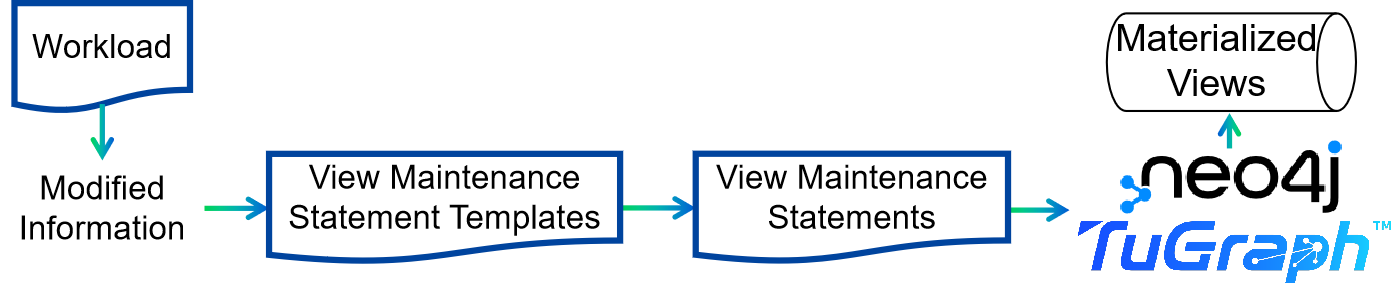}
    \caption{Overview of View Maintenance}
    \label{fig:view_maintenance}
\end{figure}

After the view is created, any write operations such as creating, deleting, or property updating in a graph database require updating the view to maintain consistency. As shown in ~\cref{fig:view_maintenance}, we predefine maintenance statement templates for views. When the graph database is modified, these templates are iterated over,  with the relevant parameters replaced, and the resulting statements are executed in the graph database system. 

Since the views in this paper do not involve properties, we focus solely on creation and deletion operations. Specifically, we handle four cases: creating a node, deleting a node, creating an edge, and deleting an edge. Next, we will explain how to generate maintenance statement templates for these scenarios.

\mysect{Create a node.} Since simply creating a node, which is not connected to any other node, does not affect the path-based view described in this paper, it does not need to be processed. 

\mysect{Delete a node.} When deleting a node in the graph database, we need to delete all the view edges associated with that node, as shown in the ~\autoref{algo:delete_node_maintenance}. 
\begin{algorithm}[htbp]
    \caption{View Maintenance Template for Deleting a Node}
    \label{algo:delete_node_maintenance}
    \begin{algorithmic}[1]
        \REQUIRE Deleted Node Information \NF, View Creation Statements \VCS
        \ENSURE View Maintenance Template map \VMTMap
        \STATE Initialize \VMTMap as an empty map
        \FOR{\VC in \VCS}
            \STATE Initialize \VMTSet as an empty set
            \FOR{$node$ in \VC.Match.nodes}  
                \STATE \VMTSet.insert(ReplNode($node$, \NF))
            \ENDFOR
            \FOR{$e$ in \VC.Match.VLEdges}
                \STATE $n \gets e$.minLength
                \STATE $m \gets e$.maxLength
                \IF{$m == \infty$}
                    \FOR{$i=1$ to max$(n-1, 1)$}
                        \IF{$i <$ max$(n-1, 1)$}
                            \STATE \VM $\gets$ ReplNodeInVlen($(i,i), (n-i,\infty), e$, \NF)
                            \STATE \VMTSet.insert(\VM)
                        \ELSE
                            \STATE \VM $\gets$ ReplNodeInVlen(($i$,$\infty$), (1, $\infty$), e, \NF)
                            \STATE \VMTSet.insert(\VM)
                        \ENDIF
                    \ENDFOR
                \ELSE
                    \FOR{$i=1$ to $m-1$}
                        \STATE \VM $\gets$ ReplNodeInVlen(
                        \\\qquad\quad($i,i$), (max($n-i$,1), $m-i$), $e$, \NF)
                        \STATE \VMTSet.insert(\VM)
                    \ENDFOR
                \ENDIF
            \ENDFOR
            \STATE \VMTMap[\VC.\kw{ViewName}] $\gets$ \VMTSet
        \ENDFOR
    \end{algorithmic}
\end{algorithm}

In the path of a view, nodes may appear in two places, one is explicit nodes and the other is implicitly in variable-length edges. Lines 4-6 of the ~\autoref{algo:delete_node_maintenance} iterates through all the explicit nodes and calls ReplNode to generate a maintenance statement, which instantiates the node to the deleted node \NF and leaves the rest of the view unchanged, which finds all view edges when \NF is at the certain position when all view edges are found and deleted; Lines 7-26 
deal with the case in variable-length edges, setting n,m as the minimum and maximum hops of variable-length edges, respectively. We need to traverse all possible locations of the \NF, so we traverse from the smallest to the largest distance of \NF from the starting node of the variable-length edge, and we don't need to consider the case where the distance from the start node or the end node is 0 here, because a distance of 0 is an overlap with the start node or the end node, which is the case in lines 4-6 of the ~\autoref{algo:delete_node_maintenance}.

Because the maximum hop may equal $\infty$, we discuss in two cases, if the maximum hop is infinite, the only limit is the number of hops is greater than or equal to n. When the distance between \NF and the start node of \kn{e} is $i<n-1$, the distance between \NF and the end node of \kn{e} must be $>= n-i$, such as lines 13-14; 
when the distance between \NF and the start node of \kn{e} is $i>=n-1$, then the distance between \NF and the end node of \kn{e} can be $>= 1$, so this case can be merged into one statement, such as lines 16-17, 
there will be a total of $n-2+1=n-1$ statements. Of course, if $n<2$, there will also be 1 statement. 

If the maximum hop is not infinite, then traverse the distance from \NF to the start node from $1$ to $m-1$, generating a total of m-1 statements, as shown in lines 21-23. 
When the distance from \NF to the start node of \kn{e} is i, the distance from \NF to the end node of \kn{e} will be $max(n-i,1)$ at minimum, and the maximum will be $m-i$. 

Here the ReplNodeInVlen function expands the variable-length edge \kn{e} into three parts: edge-node-edge. The first parameter $(i,i)$ of ReplNodeInVlen, i.e., the minimum and maximum hops of the first edge, denotes the distance between the \NF and the start node of \kn{e}, the second parameter is the minimum and maximum hops of the second edge, which denotes the distance between the \NF and the end node of \kn{e}, and the intermediate node is instantiated with the \NF.


\mysect{Create an edge or Delete an edge.} When creating or deleting an edge, the overall idea is similar to deleting a node, creating or deleting all view edges associated with that edge, as shown in ~\autoref{algo:update_edge_maintenance}.

\begin{algorithm}[htbp]
    \caption{View Maintenance Template for Creating or Deleting an Edge}
    \label{algo:update_edge_maintenance}
    \begin{algorithmic}[1]
        \REQUIRE Modified Edge Information \EF, View Creation Statements \VCS, isCreate
        \ENSURE View Maintenance Template map \VMTMap
        \STATE Initialize \VMTMap as an empty map
        \FOR{\VC in \VCS}
            \STATE Initialize \VMTSet as an empty set
            \FOR{e in \VC.Match.FLEdges} 
                \STATE \VMTSet.insert(ReplEdge(e,\EF))
            \ENDFOR
            \FOR{e in \VC.Match.VLEdges}
                \STATE $n \gets e$.minLength
                \STATE $m \gets e$.maxLength
                \IF{$m == \infty$}
                    \FOR{$i=0$ to max$(n-1, 0)$}
                        \IF{$i <$ max$(n-1, 0)$}
                            \STATE \VM $\gets$ ReplEdgeInVlen(
                            \\\qquad\quad$(i,i), (n-1-i,\infty), e$, \EF, isCreate)
                            \STATE \VMTSet.insert(\VM)
                        \ELSE
                            \STATE \VM $\gets$ ReplEdgeInVlen(\\\qquad\quad(i,$\infty$), (0,$\infty$), $e$, \EF, isCreate)
                            \STATE \VMTSet.insert(\VM)
                        \ENDIF
                    \ENDFOR
                \ELSE
                    \FOR{$i=0$ to $m-1$}
                        \STATE \VM $\gets$ ReplEdgeInVlen(
                        \\\qquad\quad($i,i$), (max$(n-1-i,0),m-1-i$), \\\qquad\quad $e$, \EF, isCreate)
                        \STATE \VMTSet.insert(\VM)
                    \ENDFOR
                \ENDIF
            \ENDFOR
            \STATE \VMTMap[\VC.\kw{ViewName}] $\gets$ \VMTSet
        \ENDFOR
    \end{algorithmic}
\end{algorithm}

The maintenance statement for creating or deleting an edge is similar to that for deleting a node; first, all explicit edges are found and instantiated, and the replacement will instantiate the start node, the end node, and the edge itself all in \EF. 
Then deal with the case in the variable-length edge \kn{e}, also divided into two cases according to whether the maximum hop is infinite or not like ~\autoref{algo:delete_node_maintenance}, but changing $i$ to the distance from the starting node of \kn{e} to the starting node in \EF, the specific process will not be repeated.
The first parameter of ReplEdgeInVlen is the minimum and maximum number of hops for variable-length edges between the start node in \EF and the start node of \kn{e}. The second parameter is the minimum and maximum number of hops for variable-length edges between the end node in \EF and the end node of \kn{e}. The final addition of an extra \kn{isCreate} compared to ~\autoref{algo:delete_node_maintenance} to indicate whether to create an edge or to delete an edge does not affect the core of the maintenance statement, which will be explained in the subsequent examples.

\mysect{Example for View Maintenance.} 
\begin{lstlisting}[style=cypher, caption=View Example for Maintenance, label=lst:view_maintenance_example]
CREATE VIEW INDIRECT_KNOW AS (
    CONSTRUCT (s)-[r:INDIRECT_KNOW]->(d)
    MATCH (s:Person)-[:knows*3..]->(d:Person) 
)
\end{lstlisting}
For the view created by ~\autoref{lst:view_maintenance_example}, we generate templates for three view maintenance sets based on ~\autoref{algo:delete_node_maintenance} and ~\autoref{algo:update_edge_maintenance}, representing deleting a node, creating an edge, and deleting an edge, respectively.

\begin{lstlisting}[style=cypher, caption=Maintenance template for Deleting a node, label=lst:view_maintenance_node]
MATCH (s:Person:$L{$K:$V})-[:knows*3..]->(d:Person) 
WITH s,d MATCH (s)-[r:INDIRECT_KNOW NoDupEdge]->(d) DELETE r
,
MATCH (s:Person)-[:knows*3..]->(d:Person:$L{$K:$V}) 
WITH s,d MATCH (s)-[r:INDIRECT_KNOW NoDupEdge]->(d) DELETE r
,
MATCH (s:Person)-[:knows*1]->(:$L{$K:$V})-[:knows*2..]->(d:Person) 
WITH s,d MATCH (s)-[r:INDIRECT_KNOW NoDupEdge]->(d) DELETE r
,
MATCH (s:Person)-[:knows*2..]->(:$L{$K:$V})-[:knows*1..]->(d:Person) 
WITH s,d MATCH (s)-[r:INDIRECT_KNOW NoDupEdge ]->(d) DELETE r
\end{lstlisting}

\autoref{lst:view_maintenance_node} demonstrates the maintenance template for the view shown in ~\autoref{lst:view_maintenance_example} when deleting a node with a total of four statements,
i.e., $|$\VMTSet$|=4$. In ~\autoref{algo:delete_node_maintenance}, the first two statements are generated by lines 4-6. 
The last two statements are for the case where the deleted node is in the variable-length edge, and since the maximum number of hops is infinite, they correspond to lines 11-19 in ~\autoref{algo:delete_node_maintenance}.
In the example, $n=3, m=\infty$, $i = 1 \ to \ 2$. When $i$ is $1$, corresponding to lines 12-14 of ~\autoref{lst:view_maintenance_node}, the third statement of ~\autoref{lst:view_maintenance_node} is generated, i.e., the deleted node is greater than or equal to $2$ from the end node when the distance from the start node is $1$; 
When $i$ is $2$, corresponding to lines 15-18 of ~\autoref{algo:delete_node_maintenance}, when the deleted node is at a distance greater than or equal to $2$ from the start node, it is all at a distance greater than or equal to $1$ from the end node, so they can be combined as a single statement, i.e., the last statement of ~\autoref{lst:view_maintenance_node}.

It can be noted that each statement has an identical clause headed by the \kn{WITH} keyword, which is the part that removes the view edges from each pair of views found for their start and end nodes. 
One thing to note here is that for each pair of (\kn{s},\kn{d}) found, if there are multiple view edges in \kn{s} and \kn{d}, not all of them are deleted, but only one of them is deleted, because each view edge actually corresponds to one of the cases in which all the nodes and edges are instantiated in the \VPatG, and also corresponds to each pair of \kn{s},\kn{d} that we found here. 

This is correct for a single statement, since a \kn{Match} statement such as 

\ksb{MATCH (s:Person:\$L\{\$K:\$V\})-[:knows*3..]-\textgreater(d:Person)},  
\\the result of each graph instance matched by the \kn{Match} statement corresponds to a different graph instance matched in the view creation statement, so for each pair of \kn{s,d} matched, deleting or creating one of their edges is consistent with the database.

But if there are multiple statements, different maintenance statements may match to the same view, meaning that the deleted node may appear in two locations at the same time making the deleted view duplicated.
For example, if \kn{Person\{id:1\}} is deleted, the first statement of ~\autoref{lst:view_maintenance_node} may match this case: 

\ksb{MATCH (s:Person\{id:1\})-[:knows]-\textgreater(:Person\{id:2\})-[:knows]-\textgreater(d:Person\{id:1\})}.
\\ However, the second statement will match the same case, which could result in the removal of two view edges (if any) of \kn{s->d} when deleting, but only one view edge should be removed for the same instance. So we need to save the matched graph instances, i.e., the actual graph data corresponding to each node and each edge in the pattern graph of the view \VPatG, to make sure that only one view edge is deleted or created for the same graph instance.

\vspace{-2mm}
\begin{lstlisting}[style=cypher,caption=Maintenance template for Creating an edge, label=lst:view_maintenance_edge]
MATCH (s:Person)-[:knows*0]->(:$SL{$SK:$SV})-[@R:knows]->(:$DL{$DK:$DV})-[*2..]->(d:Person) WHERE id(@R)=$RID 
WITH s,d CREATE (s)-[r:INDIRECT_KNOW]->(d)
,
MATCH (s:Person)-[:knows*1]->(:$SL{$SK:$SV})-[@R:knows]->(:$DL{$DK:$DV})-[*1..]->(d:Person) WHERE id(@R)=$RID
WITH s,d CREATE (s)-[r:INDIRECT_KNOW]->(d)
,
MATCH (s:Person)-[:knows*2..]->(:$SL{$SK:$SV})-[@R:knows]->(:$DL{$DK:$DV})-[*0..]->(d:Person) WHERE id(@R)=$RID
WITH s,d CREATE (s)-[r:INDIRECT_KNOW]->(d)
\end{lstlisting}

~\autoref{lst:view_maintenance_edge} demonstrates the maintenance template for the view shown in ~\autoref{lst:view_maintenance_example} when creating an edge, by \kn{Union} combining the three statements into one, i.e., $|$\VMTSet$|=3$. All statements are for the case where the created edge is in the variable-length edge, generated by lines 11-19 of ~\autoref{algo:update_edge_maintenance}.
Deleting an edge is largely similar to creating an edge, except that the statement following the \kn{WITH} keyword is different; when deleting an edge, the portion following the \kn{WITH} keyword is the same as the maintenance template for deleting a node.
Another point of detail is that edges don't necessarily have a primary key, so we use the internal id of the graph database system to uniquely lock the edge.
%

\mysect{Correctness.} 
When creating or deleting data, the purpose of the maintenance statement is to ensure the consistency of the view and the graph database, i.e., the view after maintenance and the view after deleting the view to re-execute the create view statement need to be identical. The \kn{Match} part of our generated maintenance statement has matched all possible graph instances affected by the changed data, as long as we ensure that for each graph instance to delete/create and only delete/create a view can ensure the consistency of the view.
As previously discussed, we need to save each graph instance and just skip it the next time we match to the same graph instance, which will serve as a future work.

\mysect{Complexity Discussion.} 
Assuming the number of edges that the view maintenance statement will delete or create is $N$, the time complexity of maintenance is often $O(N)$. Although it is possible that the maintenance statement may traverse illegal nodes or edges when matching feasible paths, in fact, the more illegal nodes or edges traversed in the maintenance statement, the more traversal can be reduced by the query optimization in ~\cref{sec:query_opt1}, and the better the optimization effect of the view.

In practice, in most cases, the $N$ involved in a single node or a single edge is not large, so the maintenance time required for small-scale updates under incremental maintenance will not increase significantly, which can be verified in ~\cref{sec:evaluation}.

\section{Query Optimization Using Views}
\label{sec:query_opt1}

\begin{figure*}[]
    \centering
    \includegraphics[width=0.7\textwidth]{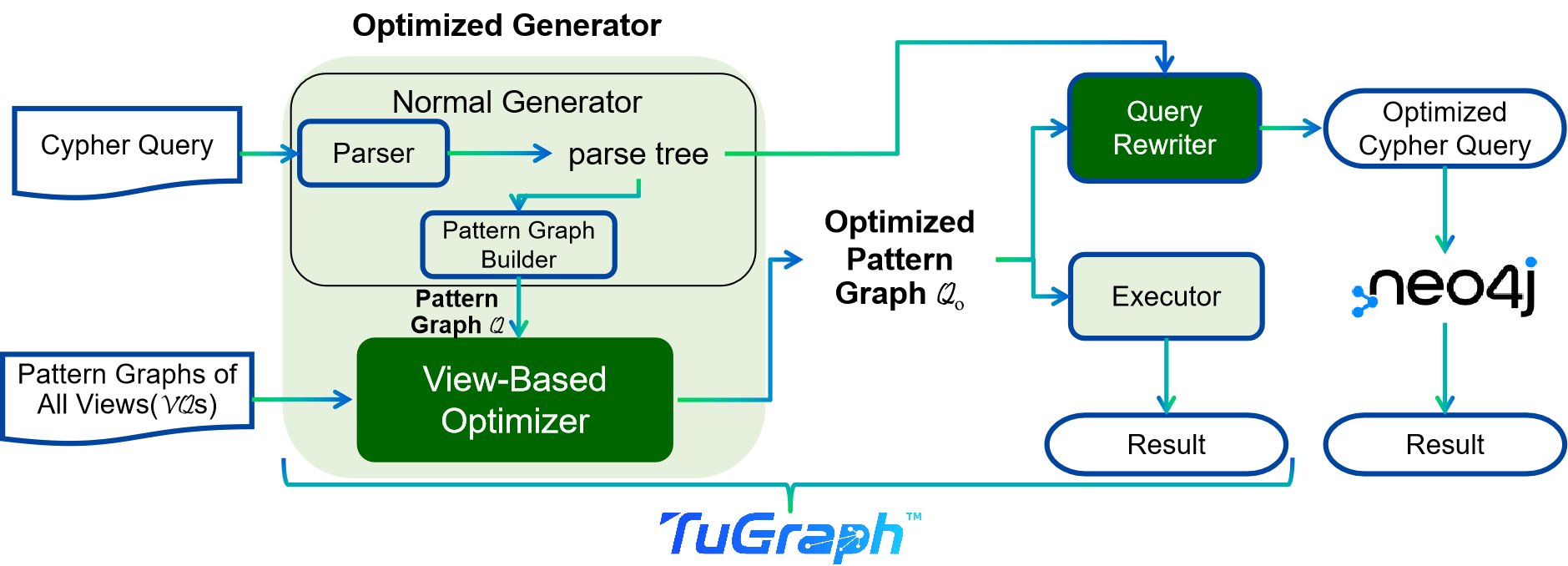}
    \caption{Overview of Query Optimization Using Views}
    \label{fig:view_optimization}
\end{figure*}

The ultimate goal of building materialized views is to optimize query efficiency. ~\cref{fig:view_optimization} shows the basic flow of {Query Optimization Using Views}, divided into two output routes, one route is executed in TuGraph for the main part in the middle, and the other route first relies on TuGraph to directly output the optimized query statement, which is then executed in Neo4j.
\begin{algorithm}[htbp]
    \caption{View-Based Optimizer for Pattern Graph}
    \label{algo:view_based optimizer}
    \begin{algorithmic}[1]
        \REQUIRE pattern graph \PatG, the set of all view pattern graphs \VPatGs
        \ENSURE optimized pattern graph \OPatG
        \STATE \SVPatGs $\gets$ SortByOptEff(\VPatGs)
        \FOR{\VPatG in \SVPatGs}
            \STATE Init()
            \WHILE{MatchView(\PatG, \VPatG, matchResult)}
                \STATE ChangePG(\PatG, \VPatG, matchResult)
                \STATE Reset()
            \ENDWHILE
        \ENDFOR
        \STATE \OPatG $\gets$ \PatG
    \end{algorithmic}
\end{algorithm}

At the heart of ~\cref{fig:view_optimization} is the \kw{View-Based Optimizer}, we use ~\autoref{algo:view_based optimizer} to describe its exact process. The algorithm first sorts all the views through the SortByOptEff function, then iterates through each view according to the order, continuously matches the view with the query graph and changes the query graph according to the \kn{matchResult}(The Init and Reset functions empty the \kn{matchResult} and other global variables), and finally outputs the optimized query graph \OPatG.




\subsection{Sorting by Optimization Effect}
\begin{figure}
  \begin{minipage}{.5\textwidth} 
    \centering
    \includegraphics[scale=0.35]{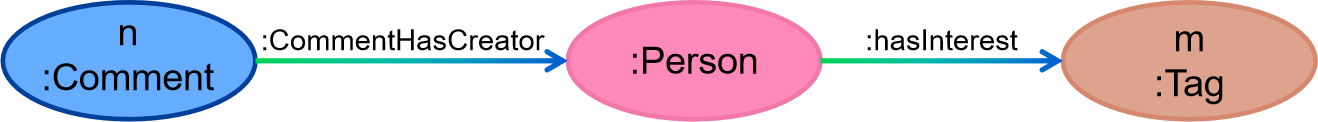}
    \caption{View1}
    \label{fig:sortView1}
  \end{minipage}

\vspace{1mm}

  \begin{minipage}{.5\textwidth} 
    \centering
    \includegraphics[scale=0.35]{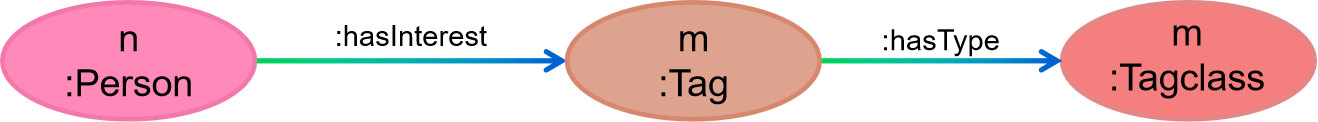}
    \caption{View2}
    \label{fig:sortView2}
  \end{minipage}
\end{figure}

\begin{figure}
  \begin{minipage}{.5\textwidth} 
    \centering
    \includegraphics[width=.95\linewidth]{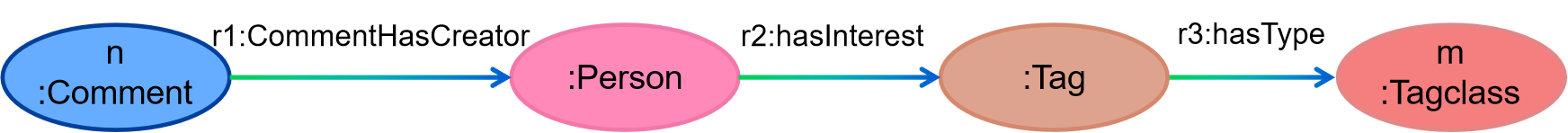}
    \caption{query pattern graph}
    \label{fig:sortPatternGraph}
  \end{minipage}

  \begin{minipage}{.5\textwidth} 
    \centering
    \includegraphics[width=.95\linewidth]{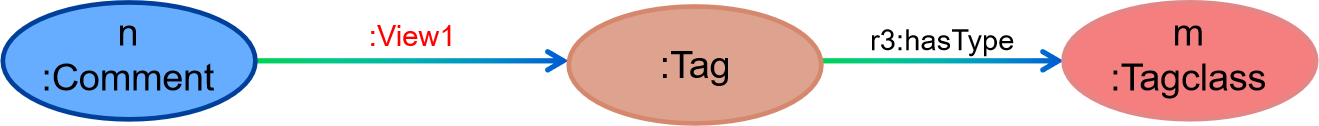}
    \caption{optimized query pattern graph 1}
    \label{fig:sortResult1}
  \end{minipage}

  \begin{minipage}{.5\textwidth} 
    \centering
    \includegraphics[width=.95\linewidth]{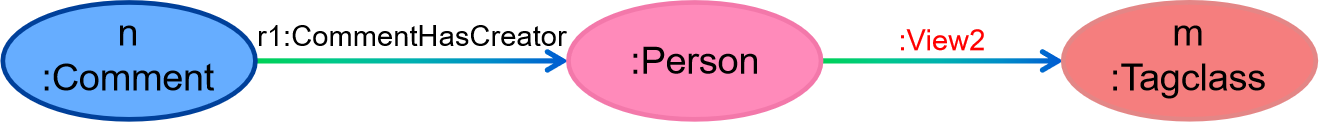}
    \caption{optimized query pattern graph 2}
    \label{fig:sortResult2}
  \end{minipage}
\end{figure}

 Different view traversal orders lead to different optimization results. For example, assuming that \VPatGs have two views in ~\cref{fig:sortView1} and ~\cref{fig:sortView2}, for the query pattern graph in ~\cref{fig:sortResult1}, if View1 is used first to optimize, then the optimized result is shown in ~\cref{fig:sortResult1}, which only applies View1; if View2 is used first to optimize, then the optimized result is shown in ~\cref{fig:sortResult2}, which only applies View2.

So the SortByOptEff function sorts the collection of views from largest to smallest according to the optimization effect of the views. Assume \DBHit{} represents the total number of accesses made by all operators to the storage engine to retrieve necessary data during the execution plan, e.g., accessing a node or an edge in the database is counted as one time. The subscript \textsf{noV} and \textsf{V} stand for ``no Views'' and ``with Views'', respectively. The currently adopted practice is to consider only the optimization effect of the view itself, i.e., \DBHit{noV} - \DBHit{V}. And \DBHit{V}=$|N_{\$SL}| + 2 * |E_{\$VL}|$($\$SL$ and $\$VL$ are the labels for the start node of view and view edge, respectively, $|N_{\$SL}|$ is the number of nodes in the graph database labeled $\$SL$, $|E_{\$VL}|$ is the number of edges in the graph database labeled $\$VL$), i.e., the cost is to traverse the starting node, then traversing all view edges adjacent to the starting node and obtaining the corresponding end node. So the final optimization effect formula is shown in ~\autoref{eq:ViewOptEff}.

\begin{equation} \label{eq:ViewOptEff}
\begin{aligned}
\text{ViewOptEff} &= \textsf{DBHit}_\textsf{noV} - \textsf{DBHit}_\textsf{V} \\
           &= \textsf{DBHit} - (|N_{\$SL}| + 2 * |E_{\$VL}|)
\end{aligned}
\end{equation}
In this example, the optimization effect of View1 is the difference between ~\cref{fig:sortPatternGraph} and ~\cref{fig:sortResult1}, where ~\cref{fig:sortPatternGraph} goes from a node of label \kn{Comment} through two hops to a node of label \kn{Tag}, which is the original query cost of View1, while ~\cref{fig:sortResult1} goes from a node of label \kn{Comment} through View1 directly to a node of label \kn{Tag}, where the query cost becomes the number of View1, and the two differences are the effect of optimization.
The $|N_{\$SL}|$, $|E_{\$VL}|$ and $\textsf{DBHit}$ will have an initial value when the view is created. Both $|N_{\$SL}|$ and $|E_{\$VL}|$ can be easily maintained by changing them when the corresponding nodes or edges are created or deleted. But the \textsf{DBHit} is more difficult to maintain, We use an estimation approach for maintenance, specifically we will save the value of initial$\textsf{DBHit}/(|initialN_{\$SL}| + 2 * |initialE_{\$VL}|)$ as the $optRate$, assuming that the ratio $optRate$ constant, then the value of $\textsf{DBHit}$ is shown in ~\autoref{eq:DB_Hit}.

\begin{equation} \label{eq:DB_Hit}
\textsf{DBHit} = (|N_{\$SL}| + 2 * |E_{\$VL}|) * optRate
\end{equation}



\begin{algorithm}[htbp]
    \caption{Matching View Algorithm}
    \label{algo:MatchView}
    \begin{algorithmic}[1]
    \REQUIRE \PatG, \VPatG, matchResult = $\emptyset$
    \ENSURE matchResult \OPatG
        \IF{matchResult.size == \VPatG.nodes.size}
            \RETURN \TRUE
        \ENDIF
        \IF{nowVN == null}
            \STATE tempVN $\gets$ nowVN
            \STATE nextVN $\gets$ \VPatG.startNode
            \FOR{n in \PatG.nodes}
                \IF{NodeCanMatch(nextVN, n)}
                    \STATE matchResult.add(nextVN, n)
                    \STATE nowVN $\gets$ nextVN
                    \IF{MatchView(\PatG, \VPatG, matchResult)}
                        \RETURN \TRUE
                    \ENDIF
                    \STATE nowVN $\gets$ tempVN
                    \STATE matchResult.delete(nextVN)
                \ENDIF
            \ENDFOR
        \ELSE
            \STATE tempVN $\gets$ nowVN
            \STATE nextVN, nextViewRelp $\gets$ GetNext(nowVN)[0]
            \FOR{n, r in GetNext(matchResult[nowVN])}
                \IF{NodeCanMatch(nextVN, n) and RelpCanMatch(nextViewRelp, r)}
                    \STATE matchResult.add(nextVN, n)
                    \STATE nowVN $\gets$ nextVN
                    \IF{MatchView(\PatG, \VPatG, matchResult)}
                        \RETURN \TRUE
                    \ENDIF
                    \STATE nowVN $\gets$ tempVN
                    \STATE matchResult.delete(nextVN)
                \ENDIF
            \ENDFOR
        \ENDIF
        \STATE \RETURN \FALSE
    \end{algorithmic}
\end{algorithm}

\subsection{Match View}
 The algorithm of MatchView is shown in ~\autoref{algo:MatchView}, the general idea is similar to the VF2 algorithm~\cite{cordella2004sub}, which is to do matching between \VPatG and \PatG from the start node of \VPatG. The \kn{matchResult} stores the matching result, \kn{nowVN} is a global variable, which indicates the latest view node that has been matched so far. When \kn{matchResult.size} and \kn{\VPatG.nodes.size} are equal, then the match is complete and returns true; conversely, if there is no feasible match, false is returned in the last line. 
Specifically, lines 4-17 iterate over all the nodes in \PatG to match the \kn{\VPatG.startNode}. If the NodeCanMatch function is true, the match result is saved and the MatchView function is called recursively, if the recursively called MatchView function finds a feasible match, then it returns true, otherwise, it backs up \kn{nowVN} to the state before the match and deletes the match result in \kn{matchResult}; 
Lines 18-32 are executed after matching the \kn{\VPatG.startNode} with \PatG, and firstly call GetNext(nowVN) function to get the unmatched neighbors of \kn{nowVN}. Since \VPatG is a path, the set size of GetNext(nowVN) is 1. Then get the set of unmatched neighbor nodes and neighbor edges of \kn{matchResult[nowVN]}(\kn{matchResult[nowVN]} is the node in \PatG that matches with nowVN), iterate through the set, and call NodeCanMatch and RelpCanMatch functions to determine whether it can match, if it can be matched, it is handled the same as the previous lines 9-15.

When the MatchView function finds a feasible match, the ChangePG function changes \PatG based on the match. For path \MPatG in \PatG that matches the view, ChangePG replaces the path between the start and end nodes in \MPatG with a view edge.

Because ChangePG replaces all nodes and edges in \MPatG other than the start and end nodes with the view edge, if there are other references to these replaced nodes or edges in \PatG (e.g., in other clauses such as \kn{WHERE}, \kn{WITH}, \kn{RETURN}, etc.), it will affect the execution of the query statement, so an attribute \kn{isReferenced} of whether or not the nodes and edges are referenced is added to each node and edge at the parser stage. 
And in the NodeCanMatch(nextVN, n) function, we need to first determine whether the label is the same, if nextVN is not the start or end node of \VPatG, then the \kn{isRenferenced} attribute of \kn{n} needs to be false, and the node of the degree of outgoing and incoming degrees add up to 2, that is, \kn{n} will have no neighbors outside \MPatG; In RelpCanMatch, we also need to make sure that the labels are the same, that there are no other references, and that the direction, min-hop and max-hop are the same.

\subsection{Analysis}
\mysect{Termination.} 
Let the number of node in the graph $G$ be $G_N$ and the number of edges be $G_E$. Every time MatchView matches and calls ChangePG to change \PatG, at least one edge becomes a view edge, so after at most $|\mathcal{Q}_E|$ matches, all the edges will become view edges, i.e., no more matches can be made, and ~\autoref{algo:view_based optimizer} must terminate.

\mysect{Complexity Analysis.}
SortByOptEff costs $|\mathbb{VQ}|*log(|\mathbb{VQ}|)$.

The worst case of MatchView iterates through all possible matches, matching node in \VPatG with every node in \PatG, i.e., $|\mathcal{Q}_N|^{|\mathcal{VQ}_N|}$, as mentioned before for each view MatchView can match successfully up to $|\mathcal{Q}_E|$ times, so time complexity is $O(\displaystyle \sum_{\mathcal{VQ} \ in \ \mathbb{VQ}}|\mathcal{Q}_E|*|\mathcal{Q}_N|^{|\mathcal{VQ}_N|})$.

ChangePG replaces the matched paths in the query graph with view edges, i.e., remove the paths first and then insert the view edges,  the complexity is $O(|\mathcal{VQ}_N|)$ for each call. The number of calls is the same as the number of successful MatchView matches, the total complexity is $\displaystyle O(\sum_{\mathcal{VQ} \ in \ \mathbb{VQ}}|\mathcal{Q}_E|*|\mathcal{VQ}_N|)$, compared to the MatchView can be ignored.

In summary, the total time complexity is $\displaystyle O(|\mathbb{VQ}|*log(|\mathbb{VQ}|) + \sum_{\mathcal{VQ} \ in \ \mathbb{VQ}}|\mathcal{Q}_E|*|\mathcal{Q}_N|^{|\mathcal{VQ}_N|})$. 
The graph schema under the property graph model is relatively simple, the number of views will not be very large, and the number of nodes and edges in \PatG and \VPatG is also very small, and the estimation of MatchView is in the worst-case scenario, in fact, due to the presence of filtering operations such as labeling for nodes, edges, etc., the real traversal number is much less than the worst-case scenario, so the time taken to optimize the query graph using the view is completely negligible compared to the actual query time, which can be verified in the ~\cref{sec:evaluation}.

\section{Evaluation}
\label{sec:evaluation}



Since view creation and view maintenance are costly, we need to evaluate the overall optimization effect after using views to prove the positive effect of MV4PG constructed in this paper. Meanwhile, to demonstrate the scalability of the approach in this paper, we have experimented on both TuGraph and Neo4j.

\subsection{Experimental Setup}
\mysect{Environment.}
We conducted experiments on a server with Ubuntu 22.04.2, two 24-core Intel Xeon Gold 6248R Processors, and 256 GB RAM. The graph databases we evaluated are Tugraph v4.1.0 and Neo4j v4.4.2. 

\mysect{Dataset.}
We use the following two data sets for evaluation:

\begin{itemize}
    \item \kb{SNB}: the social network benchmark provided by  LDBC~\cite{erling2015ldbc}, which simulates real online social networking scenarios, is a better way to evaluate the performance of a graph database in real-world scenarios. The SNB dataset used in this paper contains 3,181,724 nodes and 17,256,038 edges.
    \item \kb{Finbench}: LDBC FinBench(Financial Benchmark)~\cite{qi2023ldbc} intends to define a benchmark characterized by special data and query patterns in financial industry to test graph database systems to make the evaluation of graph databases representative, reliable and comparable, especially in financial scenarios. The Finbench dataset used in this paper contains 5,376,981 nodes and 26,633,151 edges.
\end{itemize}


\mysect{Workload.} 
We designed a small workload for each dataset to test the effect of views, each workload has 10 test statements, including 7 read statements and 3 write statements, the write statements include the creation of edges, the deletion of edges, and the deletion of nodes in three cases. For the specific content of the workload, see ~\cite{TugraphCode}.
To ensure accuracy and consistency in our experiments, each query was run five times consecutively, and the average value was taken as the execution time. For write statements, we execute a recover statement after each execution to restore the database to its initial state.
For example, if \kn{Q8} creates an edge, then the recover statement will delete the edge.

\subsection{Evaluation Results}
\mysect{View Creation.} We created three views for each dataset based on the statements in the workload, and the time spent creating the views is shown in ~\autoref{tab:view_create}.
\begin{table}
\fontsize{8}{9.5}\selectfont
\caption{View Creation Time}
\begin{center}
\begin{tabular}{c|c|c}
    \Xhline{1pt}
    View Creation Time(s)  & \textbf{TuGraph} &  \textbf{Neo4j}\\
    \Xhline{1pt}
    ROOT\_POST(SNB) & 20.79 & 35.68 \\
    \hline
    COMMENT\_PLACE(SNB) & 58.61 & 19.93 \\
    \hline
    PERSON\_PLACE(SNB) & 154.48 & 12.00 \\
    \hline
    ACCOUNT\_LOAN(Finbench) & 406.15 & 185.71 \\
    \hline
    PERSON\_COMPANY(Finbench) & 46.85 & 16.45 \\
    \hline
    COMPANY\_LOAN(Finbench) & 25.51 & 25.70 \\
    \hline
    \Xhline{1pt}
\end{tabular}
\label{tab:view_create}
\end{center}
\end{table}

\mysect{Workload Results.}

\begin{itemize}
    \item \kb{SNB:} ~\cref{fig:tugraph_snb} shows the results of SNB's workload tested in TuGraph and ~\cref{fig:neo4j_snb} shows the results of SNB's workload tested in Neo4j where ORI\_QUERY and OPT\_QUERY represent the execution time without view optimization and with view optimization, respectively, and OPT\_VPG represents the time required by the ~\autoref{algo:view_based optimizer} to change the pattern graph. And it can be seen that the OPT\_VPG is very short and negligible compared with OPT\_QUERY. The OPT\_VPG in Neo4j is longer than in TuGraph because, when executing in Neo4j, the TuGraph interface is called to get the optimized statement (as shown in~\cref{fig:view_optimization}), which introduces additional transmission time.
    
    ~\autoref{tab:view_opt_snb} summarizes the speedup ratios of SNB's workload in TuGraph and Neo4j,  where CE, DE, and DV denote creating edge, deleting edge, and deleting node, respectively. ~\autoref{tab:workload_opt_snb} counts the speedup ratios of the whole Workload, where $W_{ori}$ denotes the execution time of the unoptimized workload, $W_{opt}$ denotes the execution time of the optimized workload, and $MV$ denotes the total time for the creation of all materialized views.

     \begin{figure}
    \begin{minipage}{.5\textwidth} 
    \centering
    \includegraphics[width=.95\linewidth]{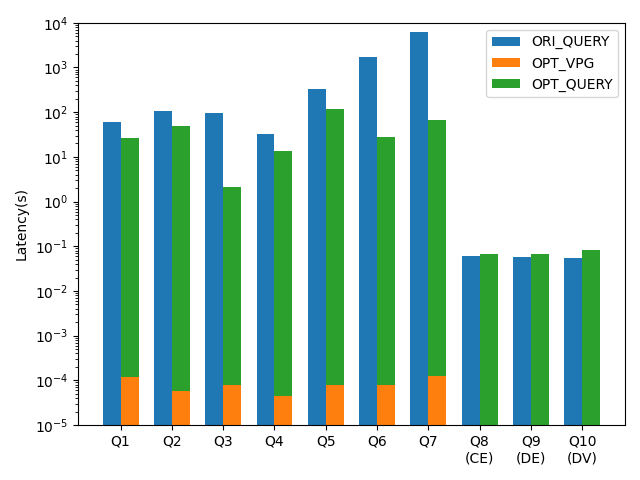}
    \caption{SNB's workload results in TuGraph}
    \label{fig:tugraph_snb}
    \end{minipage}

    \begin{minipage}{.5\textwidth} 
    \centering
    \includegraphics[width=.95\linewidth]{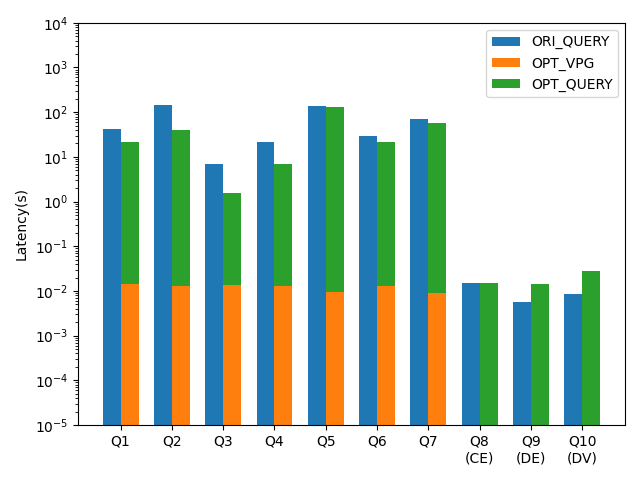}
    \caption{SNB's workload results in Neo4j}
    \label{fig:neo4j_snb}
    \end{minipage}
\end{figure}

    \begin{table}
    \fontsize{8}{9.5}\selectfont
    \caption{Speed-up ratio of SNB}
    \begin{center}
    \begin{tabular}{c|c|c}
        \Xhline{1pt}
          & \textbf{TuGraph} &  \textbf{Neo4j}\\
        \Xhline{1pt}
        Q1 & 2.27 & 1.87 \\
        \hline
        Q2 & 2.20 & 3.80 \\
        \hline
        Q3 & 45.00 & 4.41 \\
        \hline
        Q4 & 2.36 & 2.97 \\
        \hline
        Q5 & 2.81 & 1.06 \\
        \hline
        Q6 & 60.99 & 1.35 \\
        \hline
        Q7 & 96.91 & 1.27 \\
        \hline
        Q8(CE) & 0.91 & 0.99 \\
        \hline
        Q9(DE) & 0.87 & 0.40 \\
        \hline
        Q10(DV) & 0.68 & 0.30 \\
         \hline
        \Xhline{1pt}
    \end{tabular}
    \label{tab:view_opt_snb}
    \end{center}
\end{table}

    \begin{table}
    \fontsize{8}{9.5}\selectfont
    \caption{workload speed-up ratio of SNB}
    \begin{center}
    \begin{tabular}{c|c|c}
        \Xhline{1pt}
          & \textbf{TuGraph} &  \textbf{Neo4j}\\
        \Xhline{1pt}
         $W_{ori}/W_{opt}$ & 28.71 & 1.64 \\
         \hline
         $W_{ori}/(MV+W_{opt}$) & 16.19 & 1.32 \\
         \hline
        \Xhline{1pt}
    \end{tabular}
    \label{tab:workload_opt_snb}
    \end{center}

\end{table}


\begin{figure}
    \begin{minipage}{.5\textwidth} 
    \centering
    \includegraphics[width=.95\linewidth]{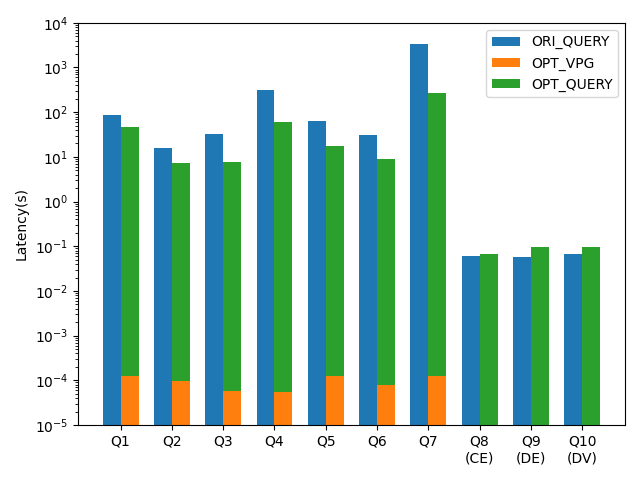}
    \caption{Finbench's workload results in TuGraph}
    \label{fig:tugraph_finbench}
    \end{minipage}

    \begin{minipage}{.5\textwidth} 
    \centering
    \includegraphics[width=.95\linewidth]{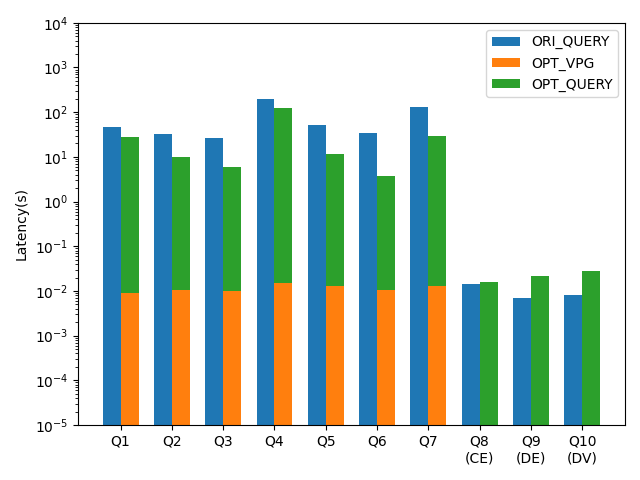}
    \caption{Finbench's workload results in Neo4j}
    \label{fig:neo4j_finbench}
    \end{minipage}
\end{figure}
 
\begin{table}
    \fontsize{8}{9.5}\selectfont
    \caption{View Opt of Finbench}
    \begin{center}
    \begin{tabular}{c|c|c}
        \Xhline{1pt}
          & \textbf{TuGraph} &  \textbf{Neo4j}\\
        \Xhline{1pt}
         Q1 & 1.94 & 1.71 \\
         \hline
         Q2 & 2.15 & 3.30 \\
         \hline
         Q3 & 4.32 & 4.42 \\
         \hline
         Q4 & 5.18 & 1.62 \\
         \hline
         Q5 & 3.54 & 4.49 \\
         \hline
         Q6 & 3.49 & 8.94 \\
         \hline
         Q7 & 12.68 & 4.34 \\
         \hline
         Q8(CE) & 0.91 & 0.91 \\
         \hline
         Q9(DE) & 0.60 & 0.34 \\
         \hline
         Q10(DV) & 0.68 & 0.29 \\
         \hline
        \Xhline{1pt}
    \end{tabular}
    \label{tab:view_opt_finbench}
    \end{center}
    
    \end{table}

\begin{table}
    \fontsize{8}{9.5}\selectfont
    \caption{workload speed-up ratio of Finbench}
    \begin{center}
    \begin{tabular}{c|c|c}
        \Xhline{1pt}
          & \textbf{TuGraph} &  \textbf{Neo4j}\\
        \Xhline{1pt}
         $W_{ori}/W_{opt}$ & 9.49 & 2.47 \\
         \hline
         $W_{ori}/(MV+W_{opt})$ & 4.43 & 1.19 \\
         \hline
        \Xhline{1pt}
    \end{tabular}
    \label{tab:workload_opt_finbench}
    \end{center}
    
    \end{table}

    \item \kb{Finbench:} ~\cref{fig:tugraph_finbench} shows the results of Finbench's workload tested in TuGraph and ~\cref{fig:neo4j_finbench} shows the results of Finbench's workload tested in Neo4j. ~\autoref{tab:view_opt_finbench} summarizes the speedup ratios of Finbench's workload in TuGraph and Neo4j, and ~\autoref{tab:workload_opt_finbench} counts the speedup ratios of the whole Workload. 
\end{itemize}


\subsection{Analysis}

\mysect{Correctness Verification.}
Comparing the results returned by each read statement, we find that they are the same. Because the view creation statement in the evaluation guarantees that there will not be a single identical node appearing multiple times in the same graph instance as previously mentioned, after executing each write statement, we find that the number of views and the number of results found in the \kn{Match} clause of the view creation statement are the same, i.e., they satisfy the data consistency, and after executing the corresponding recover operation, the number of views is also the same as the number before executing the write statement. 

\mysect{General Analysis of Performance.} 
From ~\autoref{tab:view_opt_snb} and ~\autoref{tab:view_opt_finbench}, we can see that each read statement has a relatively good acceleration ratio, and the best acceleration ratio is even close to 100x in TuGraph. For write statements, the acceleration ratios are all close to or over 30\%, which is acceptable given the very short absolute time for small updates, especially when compared to the optimization benefits achieved for read queries.

From ~\autoref{tab:workload_opt_snb} and ~\autoref{tab:workload_opt_finbench}, we can see that the speedup ratio of the whole workload is also good, and the best speedup ratio in TuGraph is as high as 28.71x. There is also a good optimization effect after adding the time of view creation, that is to say the optimization effect of adding views is greater than the additional overhead that the views bring.

\mysect{Performance Analysis for Reading Statements.} 
It can be seen that the optimization of certain statements in TuGraph is particularly good, and the optimization in Neo4j is worse than TuGraph in general, which is because the Cost-Based Optimizer used in Neo4j chooses a better execution plan, resulting in a decrease in the view optimization effect(We filed an issue on the TuGraph repository~\cite{TugraphIssue} and got a positive response!). But in general, the optimization effect of views is obvious in both graph databases, from which we select a few statements to analyze the effect of view optimization. We define the following two metrics for in-depth analysis.

\begin{definition}
    \textsf{Rows} denotes the number of data rows passed between operators in the execution plan.
    \label{def:rows}
\end{definition}

Metric \textsf{Rows} indicates the size of the intermediate results when the query is executed. A decrease in the number of intermediate results will result in less data being passed between multiple operators, thereby reducing processing time and improving query performance.

\begin{definition}
    \textsf{DBHit} denotes the number of times an operator accesses the storage engine while retrieving necessary data in the execution plan.
    \label{def:cloads}
\end{definition}

Metric \textsf{DBHit} as mentioned in \label{sec:query_opt} indicates the number of times an operator needs to access the storage engine. The decrease of \textsf{DBHit} will lead to improved query performance.



 \begin{figure}
    \centering
        \begin{minipage}{0.16\textwidth}
    \subfigure[Unoptimized Q3 of SNB] 
	{
			\centering      
			\includegraphics[width=\textwidth]{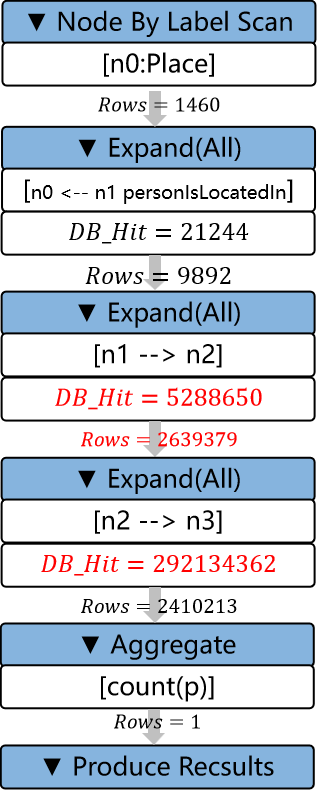}   
	}
    \end{minipage}
    \hspace{0.05\textwidth}
	\begin{minipage}{.16\textwidth}
    \subfigure[Optimized Q3 of SNB] 
	{
			\centering      
			\includegraphics[width=\textwidth]{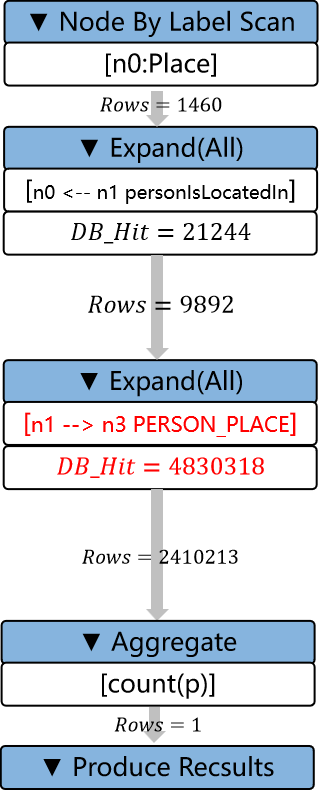}   
	}
    \end{minipage}
    \caption{Profiling data for a reading statement on TuGraph}
    \label{fig:snb_tugraph_analysis}
\end{figure}

 \begin{figure}[]
    \begin{minipage}{.32\textwidth}
    \subfigure[Unoptimized Q3 of SNB] 
	{
			\centering      
			\includegraphics[width=\textwidth]{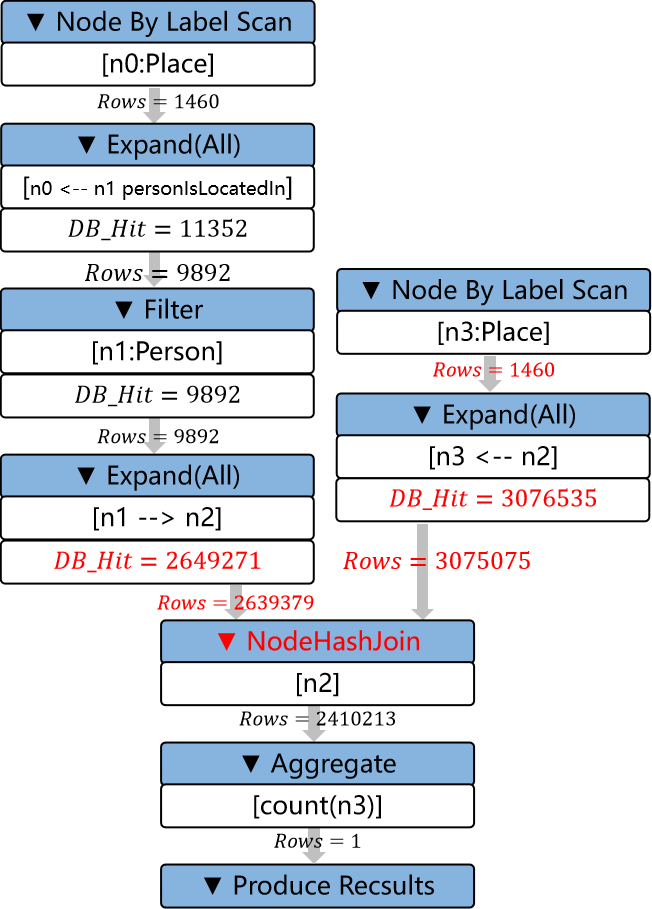}   
	}
    \end{minipage}
    \hfill
    \begin{minipage}{.16\textwidth}
    \subfigure[Optimized Q3 of SNB] 
	{
			\centering      
			\includegraphics[width=\textwidth]{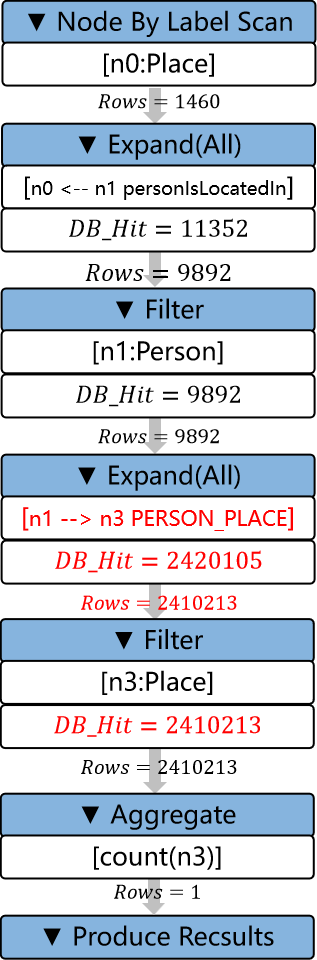}   
	}
    \end{minipage}
 \caption{Profiling data for a reading statement on Neo4j}
 \label{fig:snb_neo4j_analysis}
\end{figure}

~\cref{fig:snb_tugraph_analysis} shows the analysis results of the execution plan of Q3 in SNB on TuGraph. Q3 optimizes the two-hop path as n1-\textgreater n2-\textgreater n3 into a single edge from n1 to n3 using a view. It can be seen that the value of \textsf{DBHit} is significantly reduced, especially for the \kn{Expand} node from n2 to n3, which originally had nearly 300 million \textsf{DBHit}. However, the total \textsf{DBHit} after view optimization is only over 4 million, which is only one-sixtieth of the original. The final speedup ratio also reached 45.00.

~\cref{fig:snb_neo4j_analysis} shows the analysis results of the execution plan of Q3 in SNB on Neo4j. Without view optimization, Neo4j chooses to traverse in two paths and finally merges the two paths at n2. The optimized execution plan is similar to TuGraph. Although this results in a small difference in the number of \textsf{DBHit}, both in the millions, the extra time consumed by \kn{NodeHashJoin} still results in a speedup ratio of 4.41.

\mysect{Performance Analysis for Writing Statements.} 
\begin{figure}
    \centering
    \includegraphics[width=\linewidth]{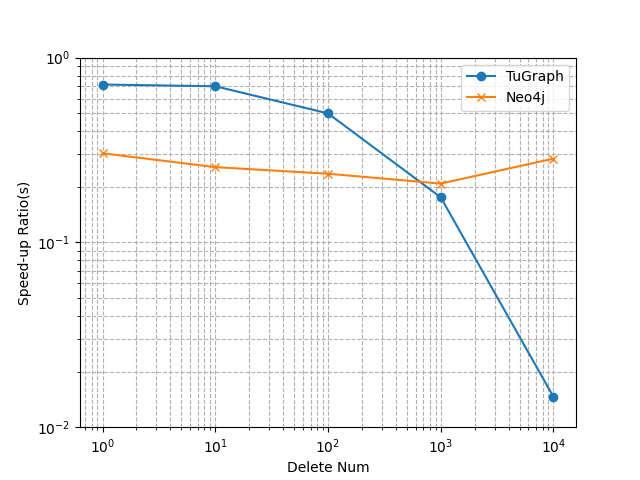}
    \caption{Update Performance}
    \label{fig:delete_performance}
\end{figure}

The efficiency of the write statements in ~\autoref{tab:view_opt_snb} and ~\autoref{tab:view_opt_finbench} is relatively high, but they only create or delete a very small number of nodes or edges. To further illustrate the speed of view maintenance, we increased the number of edges deleted to $10^{0}, 10^{1}, ..., 10^{4}$, because most updates in dynamic graphs do not involve a very large amount of data; a range of $1$ to $10^{4}$ deleted edges can cover the vast majority of cases where the number of the  total edges is on the order of tens of millions. We conducted tests in both TuGraph and Neo4j, and the speedup ratios are shown in ~\cref{fig:delete_performance}.

As we can see, the speedup ratio in TuGraph drops significantly when the number of edges reaches $10^{4}$. Our analysis indicates that this is not an issue with our incremental maintenance algorithm, which is also validated by the results from Neo4j. Instead, it's because TuGraph cannot locate a specific edge based on its edge ID. During maintenance, it must first obtain the starting node through {\$SK:\$SV } in ~\autoref{lst:view_maintenance_edge}, then traverse all adjacent edges of the starting node, and use an \kn{Edge Filter} to exclude edges that do not match the specified edge ID. When the starting node has a large number of adjacent edges, the extra traversal becomes substantial, leading to decreased efficiency. We also measured the time required by the Edge Filter during deletion in TuGraph and found that when deleting $10^{4}$ edges, the \kn{Edge Filter} accounts for 93\% of the total time. After removing the time taken by the \kn{Edge Filter}, the speedup ratios all exceed 20\%.

In Neo4j, the speedup ratio is relatively stable, remaining at 20\% to 30\% overall, and the \textsf{DBHit} and the number of deletions show a perfect linear relationship. This indicates that the overhead introduced by our maintenance algorithm is linear. In absolute time, the additional cost of view maintenance is much smaller than the efficiency gains of query optimization.
However, for graph data that is updated extremely frequently, real-time incremental maintenance may have a negative impact, which will be addressed in future work.

\section{Conclusion and Future Work}
\label{sec:conclusion}
This paper implements materialized views on property graphs and proposes for the first time a method to maintain materialized views containing variable-length edges using maintenance statement templates. Then, experiments were conducted on both TuGraph and Neo4j, and both showed excellent optimization results, demonstrating the effectiveness of the proposed views.

However, there are still some areas for improvement in current \mysys. In the view creation part, we will try to automatically select and materialize the most optimized views based on the workload and the graph schema, rather than manually selecting them as we do now. In the view maintenance part, as mentioned in ~\cref{sec:maintenance}, it is necessary to handle the special case of having the same graph instance in different statements during maintenance. This involves saving the matched graph instances and not matching them if they are the same. Additionally, we will consider seeking a balance between performance and consistency to find a better maintenance strategy, such as delayed updates. In the part of query optimization using views, the current method for selecting the view optimization order in ~\cref{sec:query_opt1} may not accurately represent the optimization effect of the views. In the future, we will consider using a Cost-Based Optimizer and a greedy algorithm to provide the query cost after applying each remaining view at each step, selecting the execution plan with the lowest cost.
All in all, graph database materialized views are still a very new field and the issues we mentioned above will be left for future work.


{\small

\bibliographystyle{IEEEtran}
\bibliography{graphdb}
}

\end{document}